# DYNAMIC OF THE ACCELERATED EXPANSION OF THE UNIVERSE IN THE DGP MODEL

VO QUOC PHONG

*The Department of Theoretical Physics, The University of Science Ho Chi Minh City*

**Abstract.** *According to experimental data of SNe Ia (Supernovae type Ia), we will discuss in detial dynamics of the DGP model and introduce a simple parametrization of matter $\omega$, in order to analyze scenarios of the expanding universe and the evolution of the scale factor. We find that the dimensionless matter density parameter at the present epoch $\Omega_m^0 = 0.3$, the age of the universe $t_0 = 12.48$ Gyr, $\frac{a}{a_0} = -2.4e^{\frac{-t}{25.56}} + 2.45$. The next we study the linear growth of matter perturbations, and we assume a definition of the growth rate, $f \equiv \frac{dln\delta}{dlna}$. As many authors for many years, we have been using a good approximation to the growth rate $f \approx \Omega_m^{\gamma(z)}$, we also find that the best fit of the growth index, $\gamma(z) \approx 0.687 - \frac{40.67}{1+e^{1.7.(4.48+z)}}$, or $\gamma(z) = 0.667 + 0.033z$ when $z \ll 1$. We also compare the age of the universe and the growth index with other models and experimental data. We can see that the DGP model describes the cosmic acceleration as well as other models that usually refers to dark energy and Cold Dark Matter (CDM).*

## I. Introduction

Over present decades, the nature of the accelerated expansion universe has been become a very special interest. A key problem that ones have to explain the origin of this phenomenon is the dark energy. The value of the dark energy density stored in the cosmological constant or a scalar field, etc. However, the cosmic expansion can be explained by extra dimensions in braneworld models without the dark energy.

The DGP model, a braneword model was proposed by Dvali, Gabadadze, Porrati in 2000, is a modified-gravity theory. The bulk of this model is an empty five-dimensional space-time with the fifth dimension which is noncompact. This theory assumed an action, which consisted the Hilbert-Einstein action in five dimensions and the Hilbert-Einstein action in 4D [2]. Thus the Friedmann equation will appear $\Omega_{r_c} = \frac{1}{4H_0^2 r_c^2}$. This appearance leads to the rate of the cosmic expansion or the age of the universe which explicitly depends on values of $\Omega_{r_c}$ and the dimensionless matter density parameter $(\Omega_m)$. i.e., $\Omega_{r_c}$ is like the dark energy density of other models and can lead to an approach about dark energy or the origin of the cosmic expansion.

Moreover, one tries to test models by experimental data, and describes the cosmic expansion by different methods. At the present, there are two ways to describe of the behavior of cosmic acceleration. The first way is to constrain between the density of pressure and the matter density. The second way rely on comparisons of the cosmic expansion history to the growth rate of Large scale structures (LSS) [43], namely determining the growth of LSS and measuring redshift distortions [44]. In this paper, we will study both above two approachs, in order to determine the age of the universe and the growth index as two tests for the DGP model.

In section II, by using SNe Ia data, we will calculate the luminosity distance of supernovae 1997ap in order to contrain $\Omega_m^0$ that is an initial parameter for calculating the age of the universe and the growth index.

In section III, we will discuss behavior of the universe and introduce a parametrization of matter that is called the matter equation of state $\omega$. Assuming $\omega$ to be constant, we will obtain the acceleration equation in order to analyze scenarios of the cosmic expansion and try to constrain the range of parameter $\omega$.

In section IV, we find that the scale factor by solving the Friedmann equation and aslo find the scale factor at early times. Then we determine the exact scenario of the cosmic expansion in the DGP model and calculating the present age of the universe $t_0$ and the transition time $t_m$ as the first test for the DGP model in section V.

Finally, we study the linear matter perturbations with the growth index $\gamma$ which is a function of redshift. At the same time, we also calculate the growth index in order to compare to experimental data as the second test for the DGP model.

## II. Luminosity distance and the value of the matter density parameter

In astronomy, an alternative way of defining a distance is through the luminosity of a stellar object [1], $d_L$ which is called the luminosity distance. The luminosity distance [1, 3, 27] is

$$d_L(z) = \frac{1+z}{H_0} \int_0^z \frac{dz'}{H(z')}, \tag{1}$$

thus we measure $d_L$ observationally, we can determine the expansion rate of the universe [1] and the dimensionless matter density parameter at the present epoch $\Omega_m^0$.

The Friedmann equation[22, 24, 25, 26, 1, 3] in DGP model that most authors have used in the following form

$$H^2 = \frac{8\pi\rho}{3m_p^2} - \frac{k}{a^2} \pm \frac{1}{r_c}\sqrt{H^2 - \frac{8\pi}{m^3}\frac{\rho_B}{6} + \frac{k}{a^2} + \frac{const}{a^4}}. \tag{2}$$

In case of the flat universe, $k = 0$, $const = 0$, bulk is empty and "-" branch [22, 24, 25, 26, 1, 3], this equation is replaced by

$$H^2 - \frac{H}{r_c} = \frac{8\pi G}{3}\rho. \tag{3}$$

The Friedmann equation (3) can be rewritten in term of the red-shift $1 + z \equiv a_0/a$ as

$$H(z)/H_0 = \sqrt{\Omega_{r_c}} + \sqrt{\Omega_{r_c} + \Omega_m^0(1+z)^3}, \tag{4}$$

where $\frac{8\pi G}{3}\rho = H_0^2 \Omega_m^0 (1+z)^3$, $\rho$ is the density of matter without the dark energy component. Here $\Omega_{r_c} = \frac{1}{4r_c^2 H_0^2}$. By using the normalization condition [26, 3] at $z = 0$ in equation (4), we get a constraint equation

$$\Omega_{r_c} = (\frac{1-\Omega_m^0}{2})^2. \tag{5}$$

Substituting (4) into (1), we obtian the luminosity distance in the DGP model,

$$d_L(z) = \frac{1+z}{H_0} \int_0^z \frac{dz'}{\sqrt{\Omega_{r_c}} + \sqrt{\Omega_{r_c} + \Omega_m^0(1+z')^3}}. \tag{6}$$

In addition to the observations of luminosity distance of high redshift supernovae, it gives rise to determining apparent magnitude of a big bang of a supernovae ($m$) which depends on the absolute magnitude ($\mathcal{M}$) and Hubble parameter $H$, is defined as [1, 3, 27, 28]:

$$m(\mathcal{M}, z) = \mathcal{M} + 5\log_{10}[H_0 d_L(z)] + 25. \tag{7}$$

According to data of 1992p, 1997ap [27], we can know $m$ of 1992p and 1997ap given in table 1,

Table 1: Data of SNe Ia

| N | $z$ | $m$ | Ref |
|---|-----|-----|-----|
| 1992p | 0.026 | 16.08 | [1, 27] |
| 1997ap | 0.83 | 24.32 | [1, 27] |

Using (1) at $z \ll 1$, we can accept $H(z)/H_0 \approx 1$, thus we can get an approximation for luminosity distance

$$d_L = \frac{1+z}{H_0} \int_0^z dz' = \frac{(1+z)z}{H_0} = \frac{z}{H_0}. \tag{8}$$

For supernovae 1992p, we can use (8), because $z_{1992p} = 0.026 \ll 1$ and according to table 1 and (7), we get $m_{1992p} = 16.0$. So we obtain

$$\mathcal{M}_{1992p} = 16.08 - 25 - 5\log_{10}^{1992p}(z/c) = -19.09 \tag{9}$$

In Astronomy, SNe Ia were used as "standard candles" and one can see that the absolute magnitudes of them are the same. Therefore according to table 1, $\mathcal{M}_{1997ap} = -19.09$, $m_{1997ap} = 24.32$. Substituting them into (7), we obtain the luminosity distance of 1997ap at $z = 0.83$

$$H_0 d_L^{1997ap} = 1.16. \tag{10}$$

Substituting (10) into (6), we find that the best fit values of $\Omega_m^0$ and $\Omega_{r_c}$ are 0.3 and 0.1225, respective. By solving (6) numeritically for the values of $\Omega_m^0$ we obtain results of luminosity distance that displayed in table 2

Table 2: $H_0 d_L$ of 1997ap

| $\Omega_m^0$ | $H_0 d_L$ |
|---|---|
| 0.25 | 1.2069 |
| 0.27 | 1.1877 |
| 0.29 | 1.1706 |
| **0.30** | **1.1627** |
| 0.31 | 1.1553 |
| 0.33 | 1.1415 |
| 0.35 | 1.1289 |

According to table 2, the best fit value of $\Omega_m^0$ is 0.3. Thus from (5) we get $\Omega_{r_c} = 0.1225$. At present by using data of Type Ia Supernovae (SNe Ia), Baryon Acoustic Oscillation (BAO), Cosmic Microwave Background (CMB), Gamma Ray Bursts (GRBs), The baryon mass fraction in clusters of galaxies (CBF), Lookback Time (LT), Growth Function (GF), Linxi Xu (2010) [28] and some authors [14, 17] also determined $\Omega_m^0 = 0.3$.

### III. Dynamic age of the universe in the DGP model

### III.1. Scenarios of the expanding universe

In the DGP model, the continuity equation[22, 24, 25, 26, 1, 3] is

$$\dot{\rho} + 3(\rho + p)H = 0, \quad (11)$$

by using this equation and equation (3) we find that the acceleration equation

$$\frac{\ddot{a}}{a} = -\frac{8\pi G(\rho+p)\left(\frac{1}{2r_c} + \sqrt{\frac{1}{4r_c^2} + \frac{8\pi G}{3}\rho}\right)}{2\sqrt{\frac{1}{4r_c^2} + \frac{8\pi G}{3}\rho}} + \left(\frac{1}{2r_c} + \sqrt{\frac{1}{4r_c^2} + \frac{8\pi G}{3}\rho}\right)^2. \quad (12)$$

We note $\omega = \frac{p}{\rho}$, $\rho \equiv \rho_m$ and $p \equiv p_m$, the acceleration equation can be rewritten as

$$\frac{\ddot{a}}{a} = \left(\sqrt{\Omega_{r_c}} + X\right)^2 h(X), \quad (13)$$

where, $X = \sqrt{\Omega_{r_c} + \Omega_m^0(1+z)^3} \geq \sqrt{\Omega_{r_c} + \Omega_m^0}$, $h(X) = 1 - \frac{3(1+w)(X-\sqrt{\Omega_{r_c}})}{2X}$. Thus the behavior of cosmic acceleration depends on the sign of $h(X)$ and we obtained the following scenarios for the expansion of the universe,

- **Scenario 01**: The universe exists only in the decelerating phase when $h(X) < 0$, $X = \sqrt{\Omega_{r_c} + \Omega_m^0(1+z)^3} \geq \sqrt{\Omega_{r_c} + \Omega_m^0}$, so we obtian the condition of $\omega$

$$\omega > \frac{1}{3\Omega_m^0} - \frac{2}{3}. \quad (14)$$

This is inconsistent with current experimental data. Showing that we are living in the accelerating universe.

- **Scenario 02**: the universe has the transition from the decelerating phase to the accelerating phase when $\omega$ satisfies

$$-\frac{1}{3} < \omega < \frac{1}{3\Omega_m^0} - \frac{2}{3}. \tag{15}$$

- **Scenario 03**: The universe exists only in the accelerating phase, ie. $h(X) > 0$ and $\forall X > \sqrt{\Omega_{r_c} + \Omega_m^0}$ and we find that

$$\omega < \frac{1}{3\Omega_m^0} - \frac{2}{3}. \tag{16}$$

### III.2. The Scale factor

The scale factor $a(t)$ is the solution of (4) and has a form

$$\frac{da}{a} = H_0(\sqrt{\Omega_{r_c}} + \sqrt{\Omega_{r_c} + \Omega_m^0(a_0/a)^3})dt, \tag{17}$$

By numerical method with the boundary condition, $t = 12.48$, $\frac{a_0}{a} = 1$ (this condition will be considered in the next section), we obtian the scale factor as figure 1

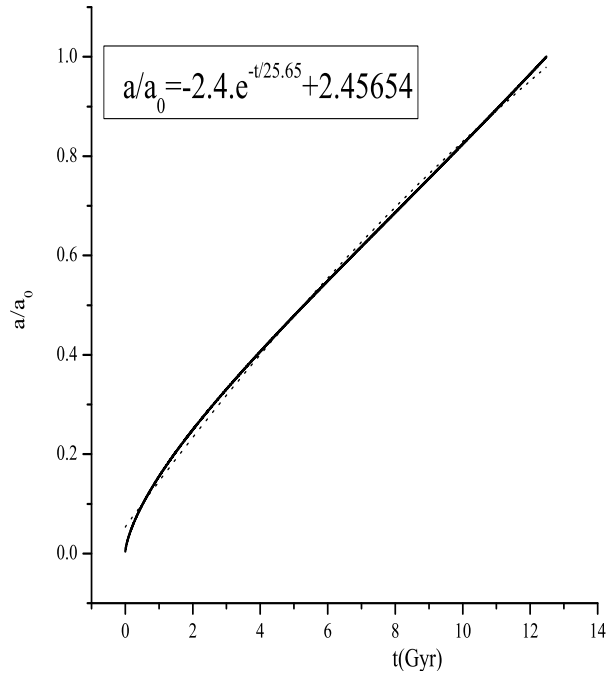

**Fig. 1.** The scale factor in the DGP model

In figure 1, the consecutive line is the graph of the numerical solution and the dashed line is an approximate graph with the following analytical expression

$$\frac{a}{a_0} = -2.4 e^{\frac{-t}{25.56}} + 2.45654. \tag{18}$$

At the time close to $t_0$ ($z$ is very small), so we can accept $H \sim H_0$ and the scale factor is given by

$$\frac{a}{a_0} = e^{H_0(t-t_0)}. \tag{19}$$

When $z \gg 1$ or $t < t_0 = 12.48$ Gyr, the equation (4) is replaced by

$$\frac{da}{a\sqrt{\Omega_m^0}(a_0/a)^{3/2}} = \frac{(\frac{a}{a_0})^{3/2} da}{a\sqrt{\Omega_m^0}} = H_0 dt. \tag{20}$$

Then the scale factor is given by

$$a/a_0 = \left(\frac{3}{2}\sqrt{\Omega_m^0} H_0 t\right)^{2/3}. \tag{21}$$

Therefore, $a \sim t^{2/3}$ when $t \ll t_0$, so we obtian

$$\dot{a} = a_0 \frac{2}{3} \left(\frac{3}{2}\sqrt{\Omega_m^0} H_0\right)^{2/3} t^{-1/3}. \tag{22}$$

We see that $\dot{a} > 0$, so the universe is always expanding. We continue to take the derivative of the above equation,

$$\ddot{a} = -a_0 \frac{2}{9} \left(\frac{3}{2}\sqrt{\Omega_m^0} H_0\right)^{2/3} t^{-4/3}. \tag{23}$$

We see that $\ddot{a} < 0$, so the universe will have a period in the deceleration phase or the universe will have the transition time $t_m$ (corresponding to $z_m$). Thus the behavior of the universe expansion must occur under scenario 2, ie $-\frac{1}{3} < \omega < \frac{1}{3\Omega_m^0} - \frac{2}{3}$,

### III.3. The age of the universe in the DGP model

The age of the universe is a major concern of astronomers and cosmologists because dynamical age of universe depends upon the rate of the expansion of the universe [5]. For many years, many groups [4, 5, 6, 7, 8, 9, 10, 11, 12, 13, 14, 15, 16, 7, 18, 19, 20, 21, 22, 23, 24, 25, 26] have calculated the present age of the universe in order to verify the model through experimental data.

In the DGP model, equation (4) yields the age of the universe in the standard form

$$t_0 H_0 = \int_0^\infty \frac{dz}{(1+z)\left[\sqrt{\Omega_{rc}} + \sqrt{\Omega_{rc} + \Omega_m^0(1+z)^3}\right]} \tag{24}$$

we can calculate $t_0$ as following when $z_s \gg 1$

$$t_0 H_0 = \int_0^{z_s} \frac{dz}{(1+z)\left[\sqrt{\Omega_{r_c}} + \sqrt{\Omega_{r_c} + \Omega_m^0(1+z)^3}\right]} + \lim_{a \to \infty}\left[\int_{z_s}^{a} \frac{dz}{\sqrt{\Omega_m^0}(1+z)^{5/2}}\right]$$
$$= \int_0^{z_s} \frac{dz}{(1+z)\left[\sqrt{\Omega_{r_c}} + \sqrt{\Omega_{r_c} + \Omega_m^0(1+z)^3}\right]} + \frac{2}{3\sqrt{\Omega_m^0}(1+z_s)^{3/2}} \quad (25)$$

Taking into the account $\Omega_m^0 = 0.3$ and choosing $H_0^{-1} = 13.77$ Gyr [26, 10, 12, 13, 14, 20] and $z_s > 10^3$, we get $t_0 H_0 = 0.9066$ by Monte Carlo or Simpson method. Thus, the age of the universe in the DGP model is $t_0 = 0.9066 \times H^{-1}{}_0 = 12.48$ Gyr.

The age of the universe is calculated much in 4D models[4, 5, 6, 7, 8, 9, 10, 11, 12, 13, 14, 15, 16, 7, 18, 19, 20, 21, 22, 23] and limited in the case $k = 0$. The best estimate is $t_0 = 13 \pm 1.5$ Gyr [4, 8, 11, 12, 17, 20] and most authors refer to dark energy. So the age of the universe in the DGP model, 12.48 Gyr, is consistent with other models.

Having $t_0$, we can calculate $t_m$, the time at which the Universe starts the accelerated phase as following:

$$t_0 H_0 - t_m H_0 = \int_0^{z_m} \frac{dz}{(1+z)\left[\sqrt{\Omega_{r_c}} + \sqrt{\Omega_{r_c} + \Omega_m^0(1+z)^3}\right]}. \quad (26)$$

Finally, in the case of $-\frac{1}{3} < w < \frac{1}{3\Omega_m^0} - \frac{2}{3}$ we obtain the following equations.

$$\begin{cases} t_0 H_0 = \int_0^{z_s} \frac{dz}{(1+z)\left[\sqrt{\Omega_{r_c}}+\sqrt{\Omega_{r_c}+\Omega_m^0(1+z)^3}\right]} + \frac{2}{3\sqrt{\Omega_m^0}(1+z_s)^{3/2}} \\ -\frac{1}{3} < w < \frac{1}{3\Omega_m^0} - \frac{2}{3} \\ z_m = \left[\frac{8+12w}{(1+3w)^2} \frac{\Omega_{r_c}}{\Omega_m^0}\right]^{1/3} - 1 \\ t_m H_0 = t_0 H_0 - I_m \\ I_m = \int_0^{z_m} \frac{dz}{(1+z)\left[\sqrt{\Omega_{r_c}}+\sqrt{\Omega_{r_c}+\Omega_m^0(1+z)^3}\right]}, \end{cases} \quad (27)$$

With $\Omega_m^0 = 0.3$ and choosing $z_s \gg 10^3$ so $t_0 H_0$ was calculated by Monte Carlo or Simpson method and we have some values of the present age of the Universe given in Table 3

Table 3: Some values with $H_0^{-1} = 13.77$ Gyr, $\Omega_m^0 = 0.3$

| $w$ | $z_m$ | $I_m$ | $t_m H_0$ | $t_m/t_0$ | $t_0 H_0$ | $t_0$ |
|---|---|---|---|---|---|---|
| -0.0072 | 0.50 | 0.346 | 0.560 | 0.618 | | |
| -0.0276 | 0.55 | 0.369 | 0.537 | 0.592 | | |
| -0.0461 | 0.60 | 0.390 | 0.516 | 0.569 | | |
| -0.0627 | 0.65 | 0.410 | 0.496 | 0.547 | | |
| -0.0779 | 0.70 | 0.429 | 0.477 | 0.526 | | |
| -0.0916 | 0.75 | 0.446 | 0.459 | 0.507 | | |
| -0.1041 | 0.80 | 0.463 | 0.443 | 0.488 | | |
| -0.1156 | 0.85 | 0.479 | 0.427 | 0.471 | | |
| -0.1262 | 0.90 | 0.494 | 0.412 | 0.455 | | |
| -0.1359 | 0.95 | 0.508 | 0.398 | 0.439 | | |
| -0.1448 | 1.00 | 0.521 | 0.385 | 0.424 | 0.9066 | 12.48 |
| -0.1531 | 1.05 | 0.533 | 0.372 | 0.411 | | |
| -0.1608 | 1.10 | 0.545 | 0.360 | 0.398 | | |
| -0.1680 | 1.15 | 0.557 | 0.349 | 0.385 | | |
| -0.1747 | 1.20 | 0.567 | 0.338 | 0.373 | | |
| -0.1809 | 1.25 | 0.578 | 0.328 | 0.362 | | |
| -0.1867 | 1.30 | 0.587 | 0.318 | 0.351 | | |
| -0.1922 | 1.35 | 0.596 | 0.309 | 0.341 | | |

We see the age of the universe as a test for describing the accelerated expansion of the universe by the DGP model. We do not use CDM but we calculated the age of the universe which is compatible with other models. This suggests that the accelerated expansion is described by the DGP model as well as other models.

Furthermore, our calculation does not refer to dark energy, but our Friedmann equation contains $\Omega_{r_c}$ which is similar to the dark energy density. So our calculation shows that the extra dimensions and the concept of dark energy may be similar to each other in describing the acceleration of expansion.

### IV. The growth index of the DGP model

According to the redshift maps of galaxies, they are distorted by its special velocity of galaxies in the line of sight. This distortions are measured by the distortion parameter $\beta$ and we can calculate the growth rate of LSS $f$ because $\beta = \frac{f}{b}$. In the cosmological models, when we surveyed the matter pertubations, we can calculate $f$. Therefore, $f$ can be used to testing cosmological models. Recent decades, many authors [33, 34, 35, 36, 37, 38, 39, 40, 41, 42, 43, 45] have applied a good approximation to $f = \Omega_m^\gamma$ with $\gamma$ is the growth index.

In 1980, Peebles [44], who made the first approximation of $f$, $f \approx \Omega_m^{0.6}$. by the FRW model. In 1991, in FRW model with cosmological constant, Lahav [44] obtained $f \approx \Omega_m^{0.6} + \frac{\Omega_\Lambda}{70}(1 + \frac{\Omega_m}{2})$. When we study the linear matter perturbations in the DGP model, we assume the growth rate [44] $f = \frac{dln\delta}{dlna}$ which is a dimensionless quantity and an analytic function in $\Omega_m$. Thus the growth rate of LSS $f$ is defined by $\delta = \frac{\delta \rho_m}{\rho_m}$ [43, 44] that satisfies the following equation at the large scales

$$(ln\delta)'' + (ln\delta)^2 + (2 + \frac{H'}{H})(ln\delta)' = \frac{3}{2}(1 + \frac{1}{3\beta})\Omega_m, \qquad (28)$$

where $(ln\delta)' = \frac{ln\delta}{dlna}$. By using the Freidmann (3), the continuity equation and the normalization condition (5), we find that

$$\Omega'_m = \Omega_m[3\frac{-1}{1+\Omega_m}(1-\Omega_m)], \qquad (29)$$

$$(ln\delta)'' = f' = \Omega_m[3\frac{-1}{1+\Omega_m}(1-\Omega_m)]\frac{df}{d\Omega_m}. \qquad (30)$$

Finally substituting (30) into (28), we can obtian

$$\Omega_m[\frac{3}{1+\Omega_m}(\Omega_m - 1)]\frac{df}{d\Omega_m} + f^2 + [\frac{1}{2} - \frac{3}{2}\frac{1}{1+\Omega_m}(\Omega_m - 1)]f = \frac{3}{2}(1 + \frac{1}{3\beta})\Omega_m. \qquad (31)$$

As many authors, we use a good approximation of $f$, $f = \Omega_m^\gamma$ [32, 33, 34, 35, 36, 37, 38, 39, 40, 41, 42, 43, 44, 45] and $\gamma$ is a function of redshift, and from (31) we arive at the equation on $\gamma(z)$ for the flat DGP model,

$$\frac{1}{2}[(1 - \frac{3(1-\Omega_m)}{1+\Omega_m}(2\gamma - 1)] - (1+z)ln(\Omega_m)\frac{d\gamma}{dz} + \Omega_m^\gamma = \frac{3}{2}(1 + \frac{1}{3\beta})\Omega_m^{1-\gamma}, \qquad (32)$$

where,

$$(1 + \frac{1}{3\beta}) = \frac{2(1+2\Omega_m^2)}{3(1+\Omega_m^2)}. \qquad (33)$$

The equation (32) is very difficult to analysis, but we can get answers when we know the boundary conditions. An effective solution for finding the boundary conditions that we approximate the above equation in the critical values of $z$ to finding the critical values of $\gamma$.

We (like Hao wei [25] (2008), Arthur Lue [3] (2004), etc.) have analysed (32) around the value $\Omega_m = 1$ and ignoring higher-order terms and we obtained

$$\gamma = \frac{9 + \Omega_m^2 + \Omega_m}{16}. \qquad (34)$$

From (34), using $\Omega_m \longrightarrow 1$ we have $\gamma \longrightarrow \frac{11}{16}$. Thus having boundary condition, we can solve (32), and obtain values of $\gamma(z)$ at $\Omega_m^0 = 0.3$ as table 4,

Table 4: values of $\gamma(z)$

| $z$ | $\Omega_m^0$ | | | |
|---|---|---|---|---|
| $z$ | 0.2 | 0.27 | 0.3 | 0.35 |
| 10 | 0.68750 | 0.6875 | *0.68750* | 0.68750 |
| 4.0 | 0.68712 | 0.68729 | *0.68734* | 0.68740 |
| 3.50 | 0.68694 | 0.68719 | *0.68726* | 0.68734 |
| 3.09 | 0.68672 | 0.68760 | *0.68715* | 0.68726 |
| **3.00** | 0.68666 | 0.68702 | **0.68712** | 0.68724 |
| 2.61 | 0.68631 | 0.68681 | *0.68694* | 0.68711 |
| 2.60 | 0.68630 | 0.68680 | *0.68694* | 0.68711 |
| 2.41 | 0.68607 | 0.68666 | *0.68682* | 0.68703 |
| 2.40 | 0.68606 | 0.68665 | *0.68682* | 0.68702 |
| 2.00 | 0.68536 | 0.68622 | *0.68646* | 0.68676 |
| 1.60 | 0.68419 | 0.68547 | *0.68584* | 0.68630 |
| **1.40** | 0.68332 | 0.68491 | **0.68536** | 0.68594 |
| 1.00 | 0.68055 | 0.68306 | *0.68380* | 0.68476 |
| **0.77** | 0.67795 | 0.68124 | **0.68225** | 0.68356 |
| 0.75 | 0.67768 | 0.68105 | *0.68280* | 0.68343 |
| **0.55** | 0.67436 | 0.67861 | **0.67996** | 0.68176 |
| 0.50 | 0.67335 | 0.67785 | *0.67928* | 0.68122 |
| 0.40 | 0.67106 | 0.67607 | *0.67770* | 0.67994 |
| **0.35** | 0.66977 | 0.67504 | **0.67678** | 0.67918 |
| 0.30 | 0.66839 | 0.67391 | *0.67576* | 0.67834 |
| 0.20 | 0.66529 | 0.67130 | *0.67338* | 0.67633 |
| 0.16 | 0.66392 | 0.67011 | *0.67228* | 0.67539 |
| **0.15** | 0.66357 | 0.66980 | **0.67200** | 0.67514 |
| 0.14 | 0.66321 | 0.66948 | *0.67170* | 0.67489 |
| 0.13 | 0.66285 | 0.66916 | *0.67140* | 0.67463 |
| 0.12 | 0.66249 | 0.66883 | *0.67110* | 0.67436 |
| 0.11 | 0.66212 | 0.66850 | *0.67078* | 0.67409 |
| 0.10 | 0.66174 | 0.66816 | *0.67047* | 0.67381 |
| 0.09 | 0.66136 | 0.66782 | *0.67014* | 0.67353 |
| 0.08 | 0.66098 | 0.66746 | *0.66981* | 0.67324 |
| 0.07 | 0.66059 | 0.66711 | *0.66947* | 0.67294 |
| 0.06 | 0.66020 | 0.66674 | *0.66913* | 0.67263 |
| 0.05 | 0.65980 | 0.66637 | *0.66878* | 0.67232 |
| 0.04 | 0.65940 | 0.66600 | *0.66842* | 0.67200 |

From results of $\gamma(z)$ at $\Omega_m^0 = 0.3$, we obtain some values of $f$,

Values of $f$ depends on $z$.

| $z$ | $\Omega_m(z)$ | $f_{the}$ |
|---|---|---|
| **0.15** | **0.369803** | **0.512479** |
| 0.35 | 0.452137 | 0.584378 |
| 0.55 | 0.521662 | 0.642445 |
| 0.77 | 0.584924 | 0.693591 |
| 1.40 | 0.710304 | 0.791017 |
| 3.00 | 0.852499 | 0.896145 |

According to experimental data, we have

Experimantal values of $f$ [35, 43].

| $z$ | $f_{obs}$ |
|---|---|
| **0.15** | **0.51 ± 0.11** |
| 0.35 | $0.52 \longrightarrow 0.88$ |
| 0.55 | $0.57 \longrightarrow 0.96$ |
| 0.77 | $0.55 \longrightarrow 1.27$ |
| 1.40 | $0.66 \longrightarrow 1.14$ |
| 3.00 | $1.17 \longrightarrow 1.75$ |

Then we compare experimental values ($f_{obs}$) and theoretical values ($f_{the}$) of growth rates, specially at $z = 0.15$, $f_{the} = f_{obs} = 0.51$. In other cases, $f_{the}$ approximate $f_{obs}$. Also according to values of $\gamma(z)$, we can draw the following graph,

According to figure 2, when $z < 1$, $\gamma(z) \sim z$ and we approximate $\gamma(z)$ as the following equation

$$\gamma(z) \approx 0.6874 - \frac{40.6774}{1 + e^{1.70155 \cdot (4.483 + z)}}, \tag{35}$$

when $z \ll 1$, we rewrite (35) as

$$\gamma(z) \approx 0.667615 + 0.0336479 z - 0.028599 z^2 + 0.0161893 z^3 + O(z^4). \tag{36}$$

Here we ignore higher-order terms of $z$ because they are very small at $z \ll 1$. Thus (36) becomes

$$\gamma(z) = 0.667615 + 0.0336479 z. \tag{37}$$

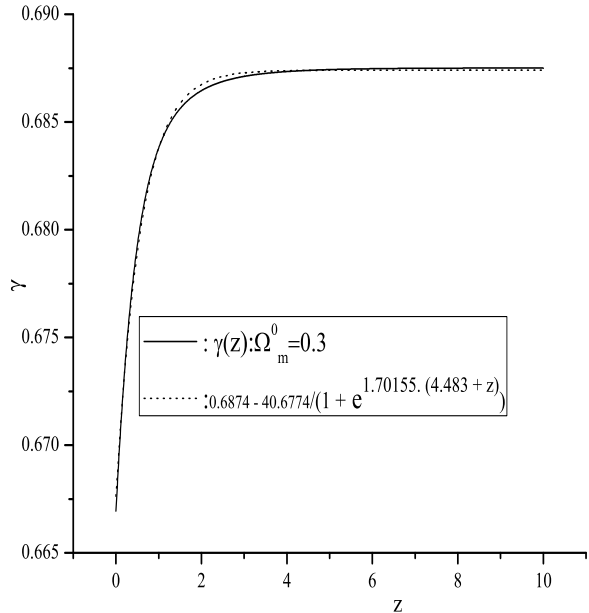

**Fig. 2.** The scale factor in the DGP model

Eq. (37) and (35) are consistent with the approximation of the current $\gamma(z)$ that is $\gamma = \gamma_0 + z.\gamma_1$ when $z < 1$ and $\gamma = \gamma_0 + \frac{z}{1+z}\gamma_1$ when $z > 1$ [33, 34, 35, 36, 37, 38, 39, 40]. Thus we show that the DGP model describes the cosmic acceleration as well as other models.

## V. Conclusion & discussion

We determined $\Omega_m^0 = 0.3$ that is the basic in order to calculate the age of the universe, 12.48 Gyr that is consistent with other models and we saw that the scenario of the cosmic expansion is the second scenario or $\frac{-1}{3} < \omega < \frac{4}{9}$. We also find the growth index and the growth rate are consistent with experimental data. Thus we can see that the DGP model describes the cosmic acceleration as well as other models that usually refers to dark energy and CDM. In addition, our calculations can be reliable basics to clarify the relationship between dark energy and extra dimensions.

However, determining the exact transition time $t_m$ in the DGP model is not possible because we do not have an experimental data or any other conditions about $\omega$ and at the present we only have experimental data of the dark energy equation of state. So, in our opinion, understanding the relationship between dark energy and extra dimensions in this model will be useful for determining $\omega$.

**Acknowledgements**

I would like to thank Professor Gia Dvali at the New York University, Professor Hoang Ngoc Long at The Institute of Physics Vietnam, Hayato Motohashi at The University of Tokyo and Dr. Vo Thanh Van at The University of Science Ho Chi Minh City for the helpful discussions about the age of the universe, SNe Ia, the growth index and the crossover scale $r_c$ in the DGP model.